\documentstyle[11pt,amssymb]{article}
\newcommand{\be}{\begin{equation}}
\newcommand{\ee}{\end{equation}}
\newcommand{\ba}{\begin{array}}
\newcommand{\ea}{\end{array}}
\newcommand{\bea}{\begin{eqnarray}}
\newcommand{\eea}{\end{eqnarray}}

\renewcommand{\l}{\newline\null}

\newcommand{\p}{\partial}
\newcommand{\ol}{\overline}

\newcommand{\la}{\langle}
\newcommand{\ra}{\rangle}

\def\hbar{h\!\!\!/}

\begin{document}
\begin{titlepage}
November 1995\hfill PAR-LPTHE 95/52
\begin{flushright} hep-th/yynnmmm \end{flushright}
\vskip 4cm
{\centerline{\bf CHIRAL SCALAR FIELDS, ELECTROWEAK
                                     {\boldmath $SU(2)_L\times U(1)$}}}
{\centerline{\bf AND THE QUANTIZATION OF THE ELECTRIC CHARGE.}}
\vskip .5cm
\centerline{B.~Machet
     \footnote[1]{Member of `Centre National de la Recherche Scientifique'.}
     \footnote[2]{E-mail: machet@lpthe.jussieu.fr.}
           }
\vskip 5mm
\centerline{{\em Laboratoire de Physique Th\'eorique et Hautes Energies,}
     \footnote[3]{LPTHE tour 16\,/\,1$^{er}\!$ \'etage,
          Universit\'e P. et M. Curie, BP 126, 4 place Jussieu,
          F 75252 PARIS CEDEX 05 (France).}
}
\centerline{\em Universit\'es Pierre et Marie Curie (Paris 6) et Denis
Diderot (Paris 7);} \centerline{\em Unit\'e associ\'ee au CNRS D0 280.}
\vskip 1.5cm
{\bf Abstract:}  I study the scalar representations of the electroweak
group of the Standard Model, with a special emphasis on their chiral
properties and on their behaviour by the discrete symmetries
$P$ and $CP$. A gauge theory for $J=0$ mesons naturally springs out,
together with a new approach to the question of the electroweak violation of
$CP$.  When acting in the 4-dimensional space of states spanned by the special
representations under scrutiny in this paper,
the electric charge appears as one of the three generators of the diagonal
$SU(2)$ and its eigenvalues are consequently quantized.
\smallskip

\vfill
\end{titlepage}
%
%
\section{Introduction.}\label{sec:INTRO}

While scalar fields play a crucial and ambiguous role in the spontaneously
broken gauge theory of electroweak processes \cite{GlashowSalamWeinberg},
pseudoscalar and scalar
mesons are traditionally attached to the chiral group of strong
interactions \cite{CurrentAlgebra}.
However, the dominance of their electroweak interactions
makes mandatory their description within the framework of a gauge theory too.
This is achieved in this paper, which also re-unites within the same
framework the fields at the origin of the breaking of the symmetry and the
observed particles (mesons).

The group $SU(2)_L\times U(1)$ of electroweak interactions has, as will be
shown below, the fundamental
property and advantage that it is a subgroup of the chiral
group $U(N)_L \times U(N)_R$ for $N$ even. Its generators can thus
be taken as $N \times N$ matrices. Similarly, the $2N^2$ Lorentz scalar and
pseudoscalar mesons fields can naturally be represented as $N \times N$
matrices. I reduce the most general $J=0$ representations of the electroweak
group to  $N^2/2$ quadruplet representations.

In the space of states spanned by these representations, the electric charge
becomes one of the three generators of the diagonal
$SU(2)$; its eigenvalues are consequently quantized.

The existence of a unique quadratic invariant for those representations also
allows to write a $SU(2)_L \times U(1)$ gauge invariant Lagrangian for $J=0$
mesons.

{From} their transformation properties by parity and charge conjugation
we can also deduce in a straightforward way non trivial results concerning
$CP$ transformations of observed electroweak states.

\section{The chiral group  \boldmath{$U(N)_L \times U(N)_R$}.}
\label{sec:CHIRAL}

A generator $\Bbb A$ of $U(N)_L \times U(N)_R$ is a set of two $N\times N$
matrices $({\Bbb A}_L, {\Bbb A}_R)$. A generator of a diagonal subgroup
satisfies ${\Bbb A}_L =  {\Bbb A}_R$.

Both left and right parts of the chiral group violate parity; hence it is
natural to classify the $J=0$ fields according to their behaviour by the
parity changing operator $\cal P$, which transforms a scalar into a
pseudoscalar and vice-versa; we shall accordingly consider the action of
the chiral group on $\cal P$-even or $\cal P$-odd states.

We define it  by the actions of its left and right commuting  subgroups.
At the level of the algebra:
\bea
{\Bbb A}^i_L\,.\, {\Bbb M}_{{\cal P}even} &\stackrel{def}{=}&
            - \,{\Bbb A}^i_L \,{\Bbb M}_{{\cal P}even}
                                           ={1\over 2} \left(
            [{\Bbb M}_{{\cal P}even},{\Bbb A}^i_L]
                            - \{{\Bbb M}_{{\cal P}even},{\Bbb A}^i_L\}
                                    \right),\cr
{\Bbb A}^i_L\,.\, {\Bbb M}_{{\cal P}odd} &\stackrel{def}{=}&
            +\,{\Bbb M}_{{\cal P}odd}\,{\Bbb A}^i_L
                                           =\hskip 5mm {1\over 2} \left(
            [{\Bbb M}_{{\cal P}odd},{\Bbb A}^i_L]
                            + \{{\Bbb M}_{{\cal P}odd},{\Bbb A}^i_L\}
                                    \right),\cr
{\Bbb A}^i_R\,.\, {\Bbb M}_{{\cal P}even} &\stackrel{def}{=}&
            +\,{\Bbb M}_{{\cal P}even} \,{\Bbb A}^i_R
                                           ={1\over 2} \left(
            [{\Bbb M}_{{\cal P}even},{\Bbb A}^i_R]
                            + \{{\Bbb M}_{{\cal P}even},{\Bbb A}^i_R\}
                                    \right),\cr
{\Bbb A}^i_R\,.\, {\Bbb M}_{{\cal P}odd} &\stackrel{def}{=}&
            - \,{\Bbb A}^i_R \,{\Bbb M}_{{\cal P}odd}
                                          =\hskip 5mm{1\over 2} \left(
            [{\Bbb M}_{{\cal P}odd},{\Bbb A}^i_R]
                            - \{{\Bbb M}_{{\cal P}odd},{\Bbb A}^i_R\}
                                    \right),
\label{eq:actionUNa}
\eea
which is akin to left- and right- multiplying $N\times N$ matrices.

{From} eqs.~(\ref{eq:actionUNa}), we see that the diagonal $U(N)$ group acts
by commutation with the ${\Bbb M}$ matrices, whatever their behaviour by
$\cal P$;
the ${\Bbb M}$'s lie in the adjoint representation of this diagonal $U(N)$.

The expressions in terms of commutators ($[\ ,\ ]$) and anticommutators
($\{\ ,\ \}$) have been kept in eq.~(\ref{eq:actionUNa}) for the reader to
make an easy link with scalars as bound states of fermions (quarks), as
evoked in the conclusion.

At the level of the group, let ${\cal U}_L \times {\cal U}_R$ be a finite
transformation of the chiral group; we have
\bea
{\cal U}_L \times {\cal U}_R \,.\, {\Bbb M}_{{\cal P}even}  &=&
            {\cal U}_L^{-1}  \,{\Bbb M}_{{\cal P}even} \,{\cal U}_R, \cr
{\cal U}_L \times {\cal U}_R \,.\, {\Bbb M}_{{\cal P}odd}  &=&
            {\cal U}_R^{-1} \,{\Bbb M}_{{\cal P}odd} \,{\cal U}_L.
\label{eq:actionUNg}
\eea
Note that  this action only coincides with
the usual action of the chiral group in the $\sigma$-model for the
$\cal P$-even  scalars, while ``left'' and ``right'' are swapped in the
action on the $\cal P$-odd scalars.

\section{The group \boldmath{$SU(2)_L \times U(1)$}.}

The generators of the ``generic'' $SU(2)$ we take as
\be
{\Bbb T}^3 = {1\over 2}\left(\begin{array}{ccc}
                        {\Bbb I} & \vline & 0\\
                        \hline
                        0 & \vline & -{\Bbb I}           \end{array}\right),\
{\Bbb T}^+ =           \left(\begin{array}{ccc}
                        0 & \vline & {\Bbb I}\\
                        \hline
                        0 & \vline & 0           \end{array}\right),\
{\Bbb T}^- =           \left(\begin{array}{ccc}
                        0 & \vline & 0\\
                        \hline
                        {\Bbb I} & \vline & 0           \end{array}\right).
\label{eq:generic}
\ee
The ${\Bbb I}$'s in eq.~(\ref{eq:generic}) stand for $N/2 \times N/2$ unit
matrices (we require ${\Bbb T}^- = ({\Bbb T}^+)^\dagger$, such that the
unit matrices are chosen to have the same dimension).
${\Bbb T}^+$ and ${\Bbb T}^-$ are respectively $({\Bbb T}^1 +
i~{\Bbb T}^2)$ and $({\Bbb T}^1 - i~{\Bbb T}^2)$.
A ``left'' generic $SU(2)$ is defined accordingly.

The $U(1)$ of hypercharge, non-diagonal, but which commutes with $SU(2)_L$
is defined by its generator $({\Bbb Y}_L, {\Bbb Y}_R)$, with
\bea
{\Bbb Y}_L &=& {1\over 6} \;{\Bbb I},\cr
{\Bbb Y}_R  &=& \hskip 2.5mm           {\Bbb Q}_R\, ,
\label{eq:U1}
\eea
where ${\Bbb Q} = ({\Bbb Q}_L , {\Bbb Q}_R)$ is the charge operator and
$\Bbb I$ is the unit $N \times N$ matrix.

The Gell-Mann-Nishijima relation
\be
{\Bbb Y} = {\Bbb Q} - {\Bbb T}\,^3_L,
\label{eq:GMN}
\ee
to be understood as
\be
({\Bbb Y}_L,{\Bbb Y}_R)= ({\Bbb Q}_L, {\Bbb Q}_R) - ({\Bbb T}\,^3_L, 0),
\label{eq:GMNchiral}\ee
is verified, for its left and right projections, by the definitions
(\ref{eq:generic},\ref{eq:U1}) above when the charge operator $\Bbb Q$ is
diagonal with
\be
{\Bbb Q}_L = {\Bbb Q}_R  =            \left(\begin{array}{ccc}
                               {2/3} & \vline & 0\\
                               \hline
                               0 & \vline & {-1/3}
\end{array}\right).
\label{eq:charge}
\ee
The ``alignment'' of the electroweak subgroup inside the chiral group is
controlled by a unitary matrix, $({\Bbb R},{\Bbb R})$, acting diagonally,
with
\be
{\Bbb R} =             \left(\begin{array}{ccc}
                        {\Bbb I} & \vline & 0\\
                        \hline
                        0 & \vline & {\Bbb K}           \end{array}\right),
\label{eq:rotation}
\ee
where $\Bbb K$ is a $N/2 \times N/2$ unitary matrix of rotation
\cite{Cabibbo,KobayashiMaskawa}.
The ``rotated'' electroweak group is then the one with generators
\be
{\Bbb R}^\dagger {\Bbb T}\; {\Bbb R}\; ;
\label{eq:rotgroup}
\ee
in practice, this rotation only acts on the ${\Bbb T}^\pm$ generators.

\section{Quadruplet scalar representations of \boldmath{$SU(2)_L \times
U(1)$}.}

Because electroweak interactions also violate parity, the representations of
the corresponding group of symmetry mix states of different parities, `scalars'
and `pseudoscalars'.
The representations are of two types, $\cal P$-even and $\cal P$-odd,
according to their transformation properties by the parity changing operator
$\cal P$ already mentioned in section \ref{sec:CHIRAL}.

In the same way (see eq.~(\ref{eq:actionUNa})) as we wrote the action of the
chiral group on scalar fields represented by $N \times N$ matrices $\Bbb M$,
we define the action of its $SU(2)_L$ subgroup, to which we add the action of
the electric charge $\Bbb Q$ according to:
\be
{\Bbb Q}\,.\,{\Bbb M} =  [{\Bbb M},{\Bbb Q}];
\label{eq:charge1}
\ee
it acts by commutation because it is a diagonal operator (see section
\ref{sec:CHIRAL}).

We shall now build a very special type of representations of the ``generic''
$SU(2)_L \times U(1)$ group defined in eqs.~(\ref{eq:generic},\ref{eq:U1}).
We write them in the form $({\Bbb M}^0, \vec {\Bbb M})$, where the $\Bbb M$'s
are still $N \times N$ matrices;
$\vec {\Bbb M}$ stand for the sets $\{{\Bbb M}^1, {\Bbb M}^2, {\Bbb M}^3\}$ or
$\{{\Bbb M}^3, {\Bbb M}^+, {\Bbb M}^-\}$ with ${\Bbb M}^+ = ({\Bbb M}^1 + i\,
{\Bbb M}^2)/\sqrt{2}\;,\; {\Bbb M}^- = ({\Bbb M}^1 - i\, {\Bbb M}^2)/\sqrt{2}$.

Let us consider quadruplets of the form
\be
({\Bbb M}\,^0, {\Bbb M}^3, {\Bbb M}^+, {\Bbb M}^-) = \left[
 {1\over \sqrt{2}}\left(\begin{array}{ccc}
                        {\Bbb D} & \vline & 0\\
                        \hline
                        0 & \vline & {\Bbb D}           \end{array}\right),
{i\over \sqrt{2}} \left(\begin{array}{ccc}
                        {\Bbb D} & \vline & 0\\
                        \hline
                        0 & \vline & -{\Bbb D}           \end{array}\right),
 i\left(\begin{array}{ccc}
                        0 & \vline & {\Bbb D}\\
                        \hline
                        0 & \vline & 0           \end{array}\right),
 i\left(\begin{array}{ccc}
                        0 & \vline & 0\\
                        \hline
                        {\Bbb D} & \vline & 0        \end{array}\right)
             \right],
\label{eq:reps}
\ee
where $\Bbb D$ is a real $N/2 \times N/2$ matrix.

The action of $SU(2)_L \times U(1)$ on these quadruplets is defined by its
action on each of the four components, as written  in
eqs.~(\ref{eq:actionUNa},\ref{eq:charge1}).
It turns out that it can be rewritten in the form
(the Latin indices $i,j,k$ run from $1$ to $3$):
\bea
{\Bbb T}^i_L\,.\,{\Bbb M}^j_{{\cal P}even} &=& -{i\over 2}\left(
              \epsilon_{ijk} {\Bbb M}^k_{{\cal P}even} +
                           \delta_{ij} {\Bbb M}_{{\cal P}even}^0
                              \right),\cr
{\Bbb T}^i_L\,.\,{\Bbb M}_{{\cal P}even}^0 &=&
                                {i\over 2}\; {\Bbb M}_{{\cal P}even}^i;
\label{eq:actioneven}
\eea
and
\bea
{\Bbb T}^i_L\,.\,{\Bbb M}_{{\cal P}odd}^j &=& -{i\over 2}\left(
                   \epsilon_{ijk} {\Bbb M}_{{\cal P}odd}^k -
                           \delta_{ij} {\Bbb M}_{{\cal P}odd}^0
                              \right),\cr
{\Bbb T}^i_L\,.\,{\Bbb M}_{{\cal P}odd}^0 &=&
                        \hskip 5mm  -{i\over 2}\; {\Bbb M}_{{\cal P}odd}^i.
\label{eq:actionodd}
\eea
The charge operator acts indifferently on ${\cal P}$-even and ${\cal P}$-odd
matrices by:
\bea
{\Bbb Q}\,.\,{\Bbb M}\,^i &=& -i\,\epsilon_{ij3} {\Bbb M}\,^j,\cr
{\Bbb Q}\,.\,{\Bbb M}\,^0 &=& 0\,,
\label{eq:chargeaction}
\eea
and the action of the $U(1)$ generator $\Bbb Y$ follows from
eq.~(\ref{eq:GMN}).

We see that we deal now with 4-dimensional representations of $SU(2)_L
\times U(1)$.
In the basis of any such representation, the generators of the electroweak
group can be rewritten as $4 \times 4$ matrices. (This is also the case for
the generators of the diagonal  $SU(2)$ evoked above, which
will be exploited in the next section concerning the quantization of the
electric charge).

We shall restrict below to this type of representations (\ref{eq:reps}).

They decompose into ``symmetric'' representations, corresponding to
$\; \Bbb D = \,{\Bbb D}^\dagger$, and ``antisymmetric'' ones for which
$\; \Bbb D = -\,{\Bbb D}^\dagger$.

There are $N/2(N/2 +1)/2$ independent real symmetric $\Bbb D$ matrices;
hence, the sets of ``even'' and ``odd'' symmetric quadruplet representations
of the type (\ref{eq:reps}) both have dimension $N/2(N/2 +1)/2$.
Similarly, the antisymmetric ones form two sets of dimension $N/2(N/2 -1)/2$.

If $({\Bbb M}\,^0, \vec {\Bbb M})$ is a representation of the ``generic''
$SU(2)_L \times U(1)$ of eqs.~(\ref{eq:generic},\ref{eq:U1}),
$({\Bbb R}^\dagger {\Bbb M}\,^0 {\Bbb R}, {\Bbb R}^\dagger
\vec {\Bbb M} {\Bbb R})$ is a representation of the ``rotated'' group of
eq.~(\ref{eq:rotgroup}); it is called hereafter a ``rotated'' representation.

Every representation above is a reducible representation of $SU(2)_L$
and is the sum of two (complex) representations of spin $1/2$.
This makes it isomorphic to the standard scalar set of the
Glashow-Salam-Weinberg model \cite{GlashowSalamWeinberg}.

Now, if we consider the transformation properties by the diagonal $SU(2)$,
all $\vec {\Bbb M}$'s are (spin $1$) triplets, lying in the adjoint
representation, while all ${\Bbb M}^0$'s are singlets.

To ease the link with physics, let us make one more step in the reshuffling
of our quadruplets.
By summing or subtracting the 2 representations, ${\cal P}$-even and
${\cal P}$-odd, corresponding to the same set of four $\Bbb M$ matrices, one
can form 2 other representations; the action of the group rewrites, using
eqs.~(\ref{eq:actioneven},\ref{eq:actionodd}):
\bea
{\Bbb T}^i_L\,.\,({\Bbb M}_{{\cal P}even}^j + {\Bbb M}_{{\cal P}odd}^j) &=&
-{i\over 2}\left(
 \epsilon_{ijk} ({\Bbb M}_{{\cal P}even}^k +{\Bbb M}_{{\cal P}odd}^k) +
   \delta_{ij} ({\Bbb M}_{{\cal P}even}^0 - {\Bbb M}_{{\cal P}odd}^0)
                              \right),\cr
{\Bbb T}^i_L\,.\,({\Bbb M}_{{\cal P}even}^0 + {\Bbb M}_{{\cal P}odd}^0) &=&
   {i\over 2}\; ({\Bbb M}_{{\cal P}even}^i - {\Bbb M}_{{\cal P}odd}^i);
\label{eq:actioneven+odd}
\eea
\bea
{\Bbb T}^i_L\,.\,({\Bbb M}_{{\cal P}even}^j - {\Bbb M}_{{\cal P}odd}^j)
&=&
-{i\over 2}\left(
 \epsilon_{ijk} ({\Bbb M}_{{\cal P}even}^k -{\Bbb M}_{{\cal P}odd}^k) +
   \delta_{ij} ({\Bbb M}_{{\cal P}even}^0 + {\Bbb M}_{{\cal P}odd}^0)
                              \right),\cr
{\Bbb T}^i_L\,.\,({\Bbb M}_{{\cal P}even}^0 - {\Bbb M}_{{\cal P}odd}^0)
&=&
   {i\over 2}\; ({\Bbb M}_{{\cal P}even}^i + {\Bbb M}_{{\cal P}odd}^i).
\label{eq:actioneven-odd}
\eea
It is convenient to rewrite
\be
({\Bbb M}_{{\cal P}even} + {\Bbb M}_{{\cal P}odd}) = {\Bbb S},
\label{eq:scalar}
\ee
and
\be
({\Bbb M}_{{\cal P}even} - {\Bbb M}_{{\cal P}odd}) = {\Bbb P},
\label{eq:pseudo}
\ee
eq.~(\ref{eq:scalar}) corresponding to a scalar state ${\Bbb S}$, and
eq.~(\ref{eq:pseudo}) to a pseudoscalar state ${\Bbb P}$.
Thus, of those two new representations, the first is of the type
\be
({\Bbb M}\,^0, \vec {\Bbb M}) = ({\Bbb S}^0, \vec {\Bbb P})
\label{eq:SP}
\ee
and the second of the type
\be
({\Bbb M}\,^0, \vec {\Bbb M}) = ({\Bbb P}\,^0, \vec {\Bbb S});
\label{eq:PS}
\ee
both have scalar and pseudoscalar entries, each entry having a definite $P$
{\em quantum number} (we attribute to scalars the parity $P = +1$ and
to pseudoscalars the parity $P = -1$).
Among the ``symmetric'' $({\Bbb S}^0, \vec {\Bbb P})$ representations lies
the one
corresponding to ${\Bbb D} = {\Bbb I}$ and which thus includes the scalar
$U(N)$ singlet: it is hereafter identified with the Higgs boson $H$ and the
corresponding representation with the usual scalar 4-plet of the
Standard Model.

{From} now onwards, we shall work with the representations (\ref{eq:SP}) and
(\ref{eq:PS}).

By hermitian conjugation a ``symmetric'' $({\Bbb M}\,^0, \vec {\Bbb M})$
representation gives $({\Bbb M}\,^0, -\vec {\Bbb M})$; an ``antisymmetric''
representation gives $(-{\Bbb M}\,^0, \vec {\Bbb M})$; the representations
(\ref{eq:SP}) and (\ref{eq:PS}) are consequently
representations of given $CP$\ (charge\ conjugation\ $\times$\  parity):
``symmetric'' $({\Bbb S}^0, \vec {\Bbb P})$'s and ``antisymmetric''
$({\Bbb P}\,^0, \vec {\Bbb S})$'s are $CP$-even, while ``symmetric''
$({\Bbb P}\,^0, \vec {\Bbb S})$'s and ``antisymmetric''
$({\Bbb S}^0, \vec {\Bbb P})$'s are $CP$-odd.

\section{The \boldmath{$SU(2) \times U(1)$} invariant Lagrangian for
scalar fields.}

To every representation (\ref{eq:reps}) is associated a unique quadratic
form invariant by any $SU(2)_L\times U(1)$ transformation:
\be
{\cal I} = {\Bbb M}\,^0 \ast ({\Bbb M}\,^0)^\dagger +
                 \vec {\Bbb M} \ast (\vec {\Bbb M})^\dagger;
\label{eq:invar}
\ee
the ``$\ast$'' product is {\em not} meant in the sense of the usual
multiplication
of matrices but in the sense of the product of fields as functions of
space-time. $\vec {\Bbb M} \ast (\vec {\Bbb M})^\dagger$ stands for
$\sum_{i=1,2,3} {\Bbb M}\,^i \ast  ({\Bbb M}\,^i)^\dagger$.

Once we have the action of the (gauge) group and a quadratic invariant, we can
immediately write a gauge invariant electroweak Lagrangian for the $2N^2$
scalar and pseudoscalar fields. It includes {\em a priori} $N^2/2$
independent mass scales.
\be
{\cal L} = \sum_{all\ reps\ {\cal R}}{1\over 2} \left(
D_\mu {\Bbb M}_{\cal R}^0 \ast (D^\mu {\Bbb M}_{\cal R}^0)^\dagger
    + D_\mu \vec {\Bbb M}_{\cal R} \ast (D^\mu \vec {\Bbb M}_{\cal R})^\dagger
- m_{\cal R}^2 ({\Bbb M}_{\cal R}^0 \ast ({\Bbb M}_{\cal R}^0)^\dagger +
                 \vec {\Bbb M}_{\cal R} \ast (\vec {\Bbb M}_{\cal R})^\dagger)
                                 \right).
\label{eq:lagrangian}
\ee
$D_\mu$ in eq.~(\ref{eq:lagrangian}) is the covariant derivative with
respect to $SU(2)_L \times U(1)$.

\section{Quantization of the electric charge.}

We have already mentioned the importance of the ``diagonal" $SU(2)$ group.
The 4-dimensional representations (\ref{eq:reps}) of $SU(2)_L \times U(1)$
being also representations of this group of symmetry,
its generators ${\Bbb T}^3, {\Bbb T}^\pm$, when acting in the 4-dimensional
vector space of which (\ref{eq:reps}) form a basis, can be represented
as $4\times 4$ matrices $\tilde T^3, \tilde T^\pm$; explicitly:
\be
\tilde{\Bbb T}^+ = \left( \ba{cccc}
                   0  &  0        &  0       &  0  \cr
                   0  &  0        & \sqrt{2} &  0  \cr
                   0  &  0        &  0       &  0  \cr
                   0  & -\sqrt{2} &  0       &  0
\ea\right),\quad
\tilde{\Bbb T}^- = \left( \ba{cccc}
                   0  &  0        &  0       &  0   \cr
                   0  &  0        &  0       & -\sqrt{2} \cr
                   0  &  \sqrt{2} &  0       &  0 \cr
                   0  &  0        &  0       &  0
\ea\right),\quad
\tilde{\Bbb T}^3 = \left( \ba{cccc}
                   0  &  0        &  0       &  0  \cr
                   0  &  0        &  0       &  0  \cr
                   0  &  0        & -1       &  0  \cr
                   0  &  0        &  0       &  1
\ea\right).
\ee
That the first line in any of the three above matrices identically vanishes
is the translation of the already mentioned fact that the first entry (${\Bbb
M}^0$) of the representations (\ref{eq:reps}) are singlets by the diagonal
$SU(2)$, while the three other entries ($\vec{\Bbb M}$) form a triplet in
the adjoint representation.

In the same way, the charge operator becomes  $\tilde Q$:
\be
\tilde{\Bbb Q} = \left( \ba{cccc}
                   0  &  0        &  0       &  0  \cr
                   0  &  0        &  0       &  0  \cr
                   0  &  0        & -1       &  0  \cr
                   0  &  0        &  0       &  1
\ea\right).
\ee
So, it occurs that $\tilde Q$ is identical with $\tilde T^3$ and that
we have the commutation relation
\be
[{\tilde T}^+, {\tilde T}^-] = 2\;{\tilde T}^3 = 2\;{\tilde Q}.
\ee
$\tilde Q$ being an $SU(2)$ generator, its eigenvalues, hence the electric
charges of the representations (\ref{eq:reps}), are quantized.

\section{Summary of the group structure.}\label{sec:GROUP}

It is instructive to summarize the group structure that we have been dealing
with.  Everything occurs ``inside'' the chiral $U(N)_L \times U(N)_R$ group.
The ``generic''
$SU(2)_L$ group is ``aligned'' with the chiral group and lies of course in
the ``left'' part of it. The ``rotated'' $SU(2)_L$ group is deduced from the
latter by a unitary matrix of rotation (Cabibbo, Kobayashi-Maskawa).
The $U(1)$ group is not diagonal, and thus extends in a non-symmetrical way
on both the ``left'' and ``right'' sides of the chiral group.
The diagonal  $SU(2)$, built from the ``rotated'' $SU(2)_L$
and its `right'' image, includes the electric charge $\tilde{\Bbb Q}$ as one
of its generators.
The latter is also, according to the Gell-Mann-Nishijima equation
(\ref{eq:GMNchiral}), the diagonal part of the above mentioned $U(1)$ group.

\section{Particles: a few brief remarks.}\label{sec:PARTICULE}

The ``standard'' scalar 4-plet of the Glashow-Salam-Weinberg model is
identified with the symmetric $({\Bbb S}^0, \vec {\Bbb P})$ representation
including the scalar $U(N)$
singlet, represented by the $N \times N$ unit matrix. The latter can
be chosen as the only diagonal $N \times N$ matrix with a non vanishing trace,
and be unambiguously defined as the Higgs boson. It is the only field
supposed to have a non-vanishing vacuum expectation value.

Any linear combination of the above representations also being a
representation, only physical observation can guide us towards the
determination of what are the observed electroweak eigenstates.
Mixing matrices link physical states with the (rotated) representations
displayed above; they can {\em a priori} depend on new parameters,
differing or not from the angles and phases characterizing, in  the rotation
matrix $\Bbb R\,$, the alignment of the electroweak group inside the
chiral group.
 Combining representations of both types $({\Bbb S}^0,\vec {\Bbb P})$ and
$({\Bbb P}\,^0, \vec {\Bbb S})$ seems however not desired since it would mix
states of different parities.

Any state produced by strong interactions is a combination of electroweak
eigenstates, which evolve and decay according to the dynamics of electroweak
interactions if no strong channel is allowed for the decay of the initial
state.

Only those representations of the $({\Bbb S}^0,\vec {\Bbb P})$ type,
and for which the
scalar entry has a non-vanishing component on the Higgs boson, will witness
leptonic decays of their three $\vec {\Bbb P}$ entries; indeed, only for those
representations will the kinetic term in the Lagrangian (\ref{eq:lagrangian})
include a
$\la H \ra \sigma^\mu \p_\mu {\Bbb P}$ coupling, where $\sigma^\mu$ is a gauge
field; the direct coupling of the latter to leptons will trigger the
leptonic decay of the pseudoscalar ${\Bbb P}$. In the same way, we deduce
that, under our hypotheses, scalar mesons never leptonically decay:
by the action of the group, a scalar is connected either to a pseudoscalar,
which is supposed to have a vanishing vacuum expectation value,
or to another scalar; but the latter is always one with a vanishing
vacuum expectation value since the Higgs boson can only be reached
by acting with the group on a pseudoscalar.

Semi-leptonic decays between states of the same parity can only occur
between the members of  the (diagonal) $SU(2)$ triplet of a given
quadruplet, since the gauge group only connects the entries of a
given representation; indeed, the kinetic term includes couplings of the type
$ {\Bbb P}_1 \sigma^\mu \p_\mu {\Bbb P}_2$, with the gauge field $\sigma_\mu$
giving leptons as before. In particular, a (diagonal) $SU(2)$ singlet like a
${\Bbb P}\,^0$ or a ${\Bbb S}^0$ never semi-leptonically decays into another
meson of the same parity.

So, the customary attribution of $CP$ quantum numbers and the presence or not
of semi-leptonic decays makes that the ``short-lived'' neutral kaon, which
is not observed to decay semi-leptonically, is most
probably the $SU(2)$``singlet'' of an ``antisymmetric''
$({\Bbb P}\,^0, \vec {\Bbb S})$ representation $(CP = +1)$, while the neutral
pion, if thought of as aligned with the corresponding ``strong''
eigenstate, and the ``long-lived'' kaon should {\em a priori} be looked for
in ``antisymmetric'' $({\Bbb S}^0, \vec {\Bbb P})$ representations $(CP = -1)$.

The simple remarks above already have for consequence that, would $N$ be equal
to $4$, the neutral electroweak pion and the long-lived neutral kaon cannot
be both pure
$CP$ eigenstates: there is indeed, in this case, only one ``antisymmetric''
$({\Bbb S}^0, \vec {\Bbb P})$ representation. In a world with $N=4$, $CP$ has
to be violated at the level of the eigenstates themselves. If $N=6$, there are
three such representations and consequently three different $\vec{\Bbb P}$
triplets can include a pure $CP$-odd neutral electroweak eigenstate.
But it is not enough to settle the case in favour of $N \geq 6$,
because, as already stated, leptonic decays
are forbidden for these ``antisymmetric'' representations, in which no entry
has been allowed to have a non-vanishing vacuum expectation value (an {\em
electroweakly} created ``antisymmetric'' pseudoscalar cannot decay into
leptons). This is in particular valid for the charged partners of the
neutral fields under concern, which are automatically given the same
electroweak
mass scale as the latter because of the uniqueness of the quadratic invariant
(only electromagnetic effects are here expected to lift the mass degeneracy
between them).
Hence, we are facing an alternative:\l
- either the vacuum structure of the theory is more complicated than has
been supposed, in particular ``antisymmetric'' scalars are allowed to also
have a non-vanishing vacuum expectation value;\l
- or the experimental evidence of leptonic decays for, say, the charged
electroweak kaons, is the hint that the physically observed
electroweak eigenstates include an admixture of the $CP$-even ``symmetric''
representation which includes the Higgs boson, {\em i.e.} they are {\em not}
pure $CP$ eigenstates.  This suggests that, whatever $N$,
$CP$ violation already occurs at the level of the eigenstates, and not
only at that of the interactions themselves  for $N\geq 6$
\cite{KobayashiMaskawa}.

\section{Conclusion. Outlook.}

We have so far built a gauge theory of mesons without any reference to their
eventual composite nature, and have nowhere mentioned the existence of
fermions other than leptons. However, the reader can easily check all the
computations and results by sandwiching the matrices $\Bbb M$ between
a $N$-vector $\Psi$ of ``quarks'' in the fundamental representation of $U(N)$,
an its conjugate $\ol\Psi$, and by introducing a $\gamma_5$ in the
definition of all ${\Bbb P}$ pseudoscalar states. The ``left'' and ``right''
generators are respectively given a $(1-\gamma_5)/2$ or a  $(1+\gamma_5)/2$
projectors when acting on the fermions, and  the laws of transformations of
the latter determine those of the mesons;
{\em all the group actions on $J=0$ fields that we have written can be
uniquely and straightforwardly deduced from the action on fermions when the
former are written as scalar or pseudoscalar diquark operators.}
Then, the occurrence of commutators and anticommutators in
eqs.~(\ref{eq:actionUNa},\ref{eq:charge1}) becomes natural.

In particular, by taking $\Bbb K$ in the rotation matrix of
eq.~(\ref{eq:rotation}) equal to the Cabibbo matrix \cite{Cabibbo},
the reader will recover the eight electroweak
representations exhibited in ref.~\cite{Machet3}. The Cabibbo angle is also
taken there to control the linear combinations of the (rotated)
representations which form the physical electroweak eigenstates (the
alignment between ``strong'' and ``electroweak'' charged eigenstates
supposed in ref.~\cite{Machet3} makes this the more likely).

The quarks could then be thought of as convenient technical entities only,
which could be forgotten.
We have however not considered here quantum effects, like in particular the
decays of the neutral pion into two photons. It is shown in
refs.~\cite{Machet1,Machet2,Machet3} that taking into account the
compositeness of the mesons in building a quantum theory correctly
reproduces those, too. The three works just quoted start from the fermionic
point of view and perform a determination of the observed
electroweak eigenstates in terms of the above representations;
they go also further in the study of their physical interactions and decays.

\bigskip
\begin{em}
\underline {Acknowledgments}: it is a pleasure to thank my colleagues at
LPTHE, specially O. Babelon, and G. Thompson for fruitful discussions and
comments.
\end{em}
\newpage\null
\begin{em}

\end{em}

\end{document}